# BENCHMARKING THE USE OF BPM QUADRUPOLE MOMENTS TO MEASURE EMITTANCE*

M.A. Balcewicz[†], C.Y. Tan, Fermi National Accelerator Laboratory, Batavia, IL, USA

*Abstract*

For the PIP-II program, transverse emittance in the Fermilab Booster must remain well controlled at higher bunch intensities. 4-plate beam position monitors (BPMs) have a small but measurable quadrupole moment, making it possible to infer transverse emittance. By compositing many BPMs together, it becomes possible to improve the quality of the quadrupole signal. The Fermilab Booster BPM system has been used to measure these quadrupole moments in the past year and derive emittances from them. Recent benchmarks show that the derived BPM emittances show similar emittance evolution and value to IPM and Multiwire data. This approach can both supplement and complement existing non-intercepting emittance monitors in accelerators.

## INTRODUCTION

The Fermilab Booster is a rapid cycling synchrotron that accelerates particles to 8 GeV kinetic energy and injects into the Main Injector and Recycler. For the PIP-II program bunch intensities are expected to increase to $6.7 \times 10^{12}$ particles per pulse (ppp), space charge driven emittance growth at this intensity is significant. The normalized emittance acceptance of the MI-8 line is designed to be approximately 25 $\pi$·mm·mrad [1], with an additional 2″ by 2″ collimator to remove halo [2] added later.

Intercepting multiwire (MW) systems exist in the Linac transfer line injecting into Booster, MI-8 line, and Booster beam dump. The MWs in the MI-8 line are of particular note as they show good agreement with one another [3] and provide absolute measures of emittance. Within the Booster cycle, non-intercepting Ionization Profile Monitors (IPMs) are used.

Without any fitting or calibration on the IPM system, the measured emittance is significantly larger than measured in the transfer line [4, 5]. The reason for the larger emittance has been suggested to be from a combination of space charge forces widening the distribution collected on the Micro-Channel Plate (MCP) [6] along with decreased MCP sensitivity at the closed orbit center of the bunch where more ions impact [6]. Furthermore, only the vertical channel of the IPM system is deemed reliable because the horizontal beta function at the IPM is 3× smaller than the vertical, thus reducing its sensitivity.

Although work has been done to correct these effects for the IPM system [7], it has nonetheless served to be an inspiration to find other methods to continuously measure emittances within the Booster ramp in both planes. A method for inferring the emittance using BPM quadrupole moments has been derived and implemented within the Booster. This method shows promise and is not sensitive to MCP gain effects or ion drift effects. This method is then compared to and cross calibrated against other instruments. Although this method has successfully measured horizontal emittances, this proceeding will focus on the vertical plane to minimize dispersive effects.

## BPM EMITTANCE MEASUREMENT

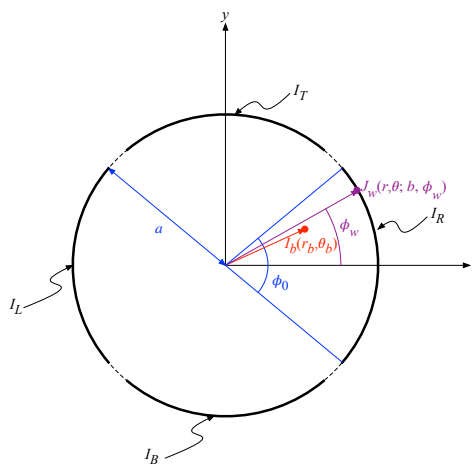

Figure 1: Transverse cross section of 4-plate BPM. Plates have a radius $a$ and subtend an angle $\phi_0$. A test current $I_b$ creates surface current $J_w$ on the plates, which can be integrated to find the total current from the test current.

For a detailed derivation for using BPMs to measure emittance please see [8, 9]. A sketch of the method will be given here for context.

Let us imagine a four plate BPM pickup with a current distribution inside it as shown in Fig. 1. The current distribution drives image currents along the BPM, the sums and ratios of which correspond primarily to the dipole moments $d_x$ and $d_y$ and total current. Plates of non-ideal geometries also measure a small quadrupole mode $q$. Assuming a given distribution, this quadrupole mode at an individual BPM can be directly related to emittance, optical functions, and momentum spread with the equation:

$$\Delta_q \equiv q - (d_x^2 - d_y^2)\phi_0 \cot \frac{\phi_0}{2} = \frac{2}{a^2} \frac{\sin \phi_0}{\phi_0} \left[ \frac{\beta_x \epsilon_x}{\pi} - \frac{\beta_y \epsilon_y}{\pi} + (D_x \sigma_p)^2 \right] \quad (1)$$

Inferring emittance in this way has been attempted in the past [10–12], but suffers from poor signal compared to the

---

large dipole signal. However, by forming a matrix equation using data from $n$ BPMs within a ring, an overdetermined set system of equations w.r.t. beam emittance is formed:

$$\begin{pmatrix} \beta_x(1) & -\beta_y(1) & D_x^2(1) \\ \beta_x(2) & -\beta_y(2) & D_x^2(2) \\ \vdots & \vdots & \vdots \\ \beta_x(j) & -\beta_y(j) & D_x^2(j) \\ \vdots & \vdots & \vdots \\ \beta_x(n) & -\beta_y(n) & D_x^2(n) \end{pmatrix} \begin{pmatrix} \epsilon_x/\pi \\ \epsilon_y/\pi \\ \sigma_p^2 \end{pmatrix} = \frac{a^2 \phi_0}{2 \sin \phi_0} \begin{pmatrix} \Delta_q(1) \\ \Delta_q(2) \\ \vdots \\ \Delta_q(j) \\ \vdots \\ \Delta_q(n) \end{pmatrix} \quad (2)$$

When approximately solved, noise from individual BPMs are suppressed in favor of the shared emittance signal seen by all BPMs. However, systematics still remain which have to be corrected, see Ref. [8].

## IPM DISTRIBUTION CORRECTION

Although issues with the IPM system were the original impetus to using the BPM system, a well calibrated IPM system is needed to evaluate the BPM emittance measurement. As such progress has been made to improve IPM measurements along the Booster ramp. To calibrate the IPM signal, one must simultaneously eliminate noise and correct for space charge effects due to ion travel. Let us begin with the IPM measurements. Booster IPMs have a turn by turn read-back, which shows rapid oscillatory noise in each channel, even when no beam is present. Figure 2 shows an FFT with three primary frequencies which vary somewhat as Booster is ramped. By adding a low pass filter, the high frequency noise is suppressed and the remaining slowly varying signal contains the signal of interest.

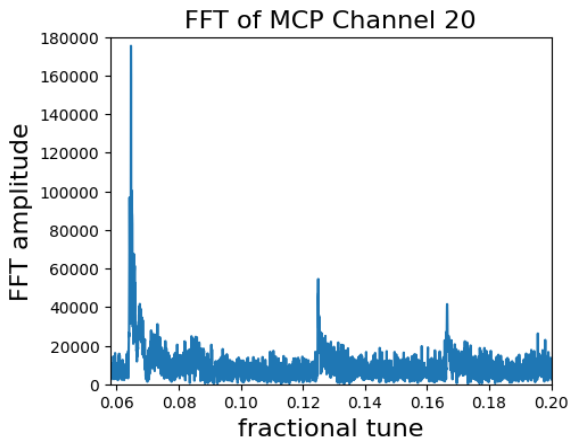

Figure 2: Primary noise spectrum in a turn by turn IPM channel near the center of the MCP. The largest of the three appears even without beam, while the other two have an intensity dependence and seem to be due to betatron oscillations around the closed orbit.

For low intensities, the IPM channels far from the bunch center still contains significant turn by turn noise within the bunch. To increase statistics, 100 revolutions are averaged together into individual data points. It is reasonable to average within 100 turns because the distribution shape, closed orbit, and optics variations are small.

By averaging the turn by turn datasets, it becomes possible to fit a distribution consistently to the IPM signals and perform background subtraction. After filtering and averaging, each IPM channel still has a small slowly varying background that can be measured with no beam. Figure 3 shows the beamless background signal for the vertical IPM. Finally, we are left with the relative sensitivity of IPM chan-nels. Areas of the MCP far from the bunch center are not excited as often, and tend to be much more sensitive to ions. This larger gain has an outsized effect on the tails of the distribution with the effect of overestimating the beam emittance. This effect is largest near injection because the geometric emittance is smallest and thus the tails have a larger contribution. Therefore, a cut at $\pm\sqrt{6}\sigma$ is made to suppress these erroneous tails.

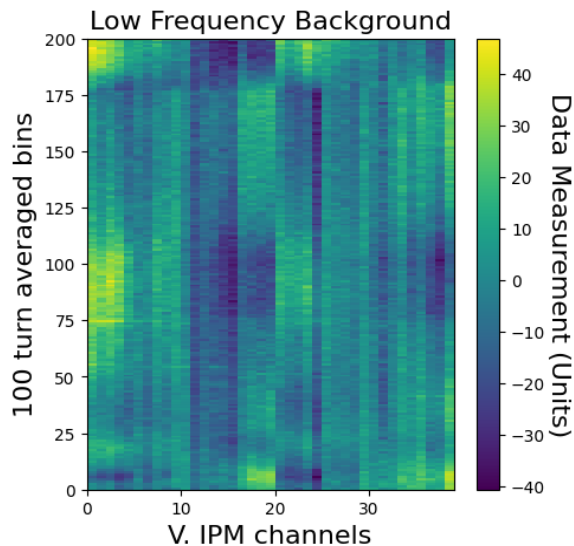

Figure 3: Background signal measured in empty beam distribution and nominal ramp. Signal has been filtered and averaged, but this does not eliminate slowly oscillating signals when could be confused with signal of interest.

## CALIBRATING IPM PARTICLE DRIFT

The IPM is a non-intercepting device and measures beam size indirectly without destroying the beam. In contrast, the multiwire systems in the MI-8 transfer line measures the beam size directly by intercepting it which destroys it in the process. It is assumed that the vertical emittance is the same right at Booster extraction and after injection into the MI-8 line. Figure 4 compares the IPM distribution to the MW distribution, which shows that by reducing the width of the IPM distribution by 2×, it can be made to match the MW distribution. The broadening of the IPM distribution comes from ion drift.

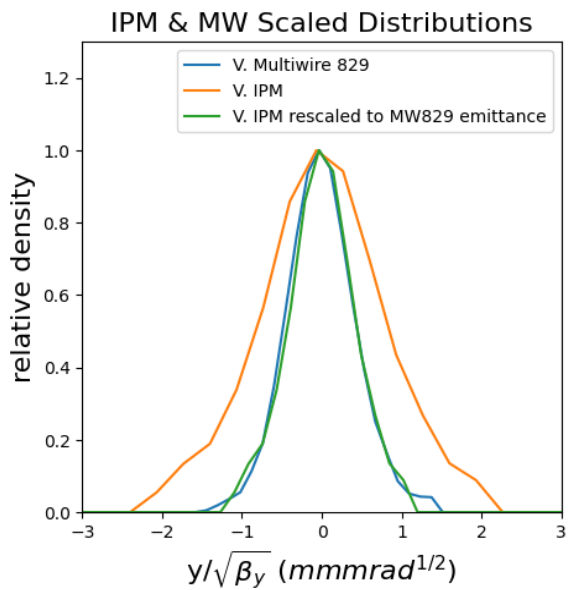

Figure 4: Shape of IPM and multiwire distributions around beam extraction. The larger measured IPM distribution is due to ion drift. The IPM shape can match the MW shape by reducing its width by 2×. IPM has suppressed tails at $\pm\sqrt{6}\sigma$ to prevent noise around tails from dominating.

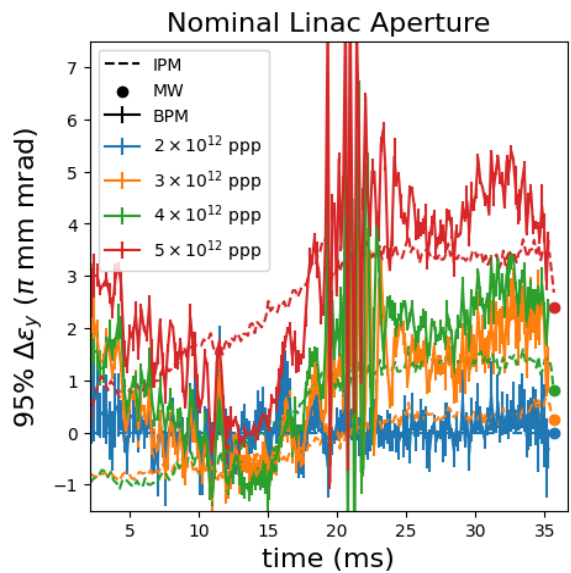

Figure 5: BPM, MW, and IPM measurements at series of intensities at nominal settings. While there is significant deviation between IPM and BPM emittance measurements, both show similar magnitude of emittance growth and qualitative behavior. BPM emittance measurement assumes the beam shape is always gaussian.

The space charge broadening of the IPM signal does not remain constant over the acceleration cycle, The broadening of the beam distribution at any point within the ramp can be destructively measured by bumping it into a collimator and measuring the surviving current. Measurements show that the low intensity ($\leq 2 \times 10^{12}$ ppp) injection emittance remains relatively constant at around 10 to 11 π·mm·mrad.

## BPM AND IPM COMPARISON

After calibrating the IPM emittance, the accuracy of the BPM quadrupole system may be evaluated. Figures 5-6 show the relative emittance measured by the methods under normal conditions and regularly pinged conditions respectively. There is some agreement between the two methods, albeit with parts of the ramp where measurements deviate significantly. This is not surprising because the BPM emittance measurement method infers changes in emittance from an assumed gaussian beam distribution. As the beam intensity is increased, the distribution deviates from gaussian and thus the derived emittance is incorrect. Although sufficient for operational needs, future optimizations of the method may improve the quality of the method. The BPM emittance measurement method is possibly applicable to any number of ringed machines that has BPMs, especially those without non-intercepting emittance instruments.

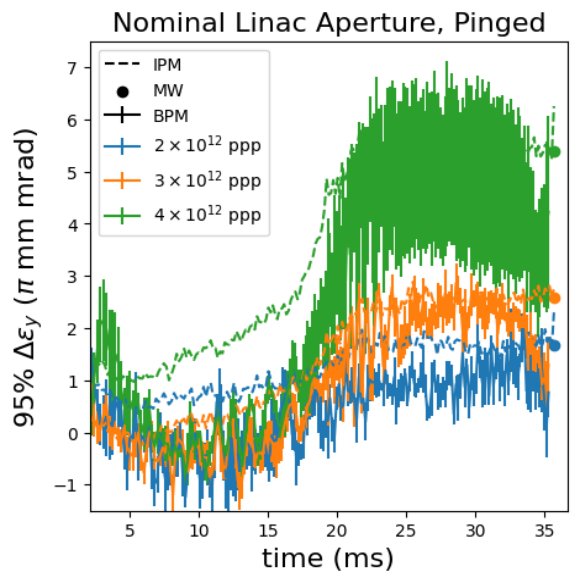

Figure 6: BPM, MW, and IPM measurements at series of intensities with repeated dipole pings. The spread in the BPM emittance comes from the pinger not kicking the beam the same every time.

## CONCLUSION

Finally, improvements in calibrating and correcting the IPM system have improved our understanding of the emittance along the Fermilab Booster ramp. Future work should be undertaken to more accurately measure beam optics and longitudinal dynamics to improve the accuracy of all of these methods and allow us to study horizontal beam distributions.